# Status of Zero Degree Calorimeter for CMS Experiment

O. A. Grachov, M. J. Murray

*Department of Physics and Astronomy, University of Kansas*
*Lawrence, KS, USA*

A. S. Ayan, P. Debbins, E. Norbeck, Y. Onel

*Department of Physics and Astronomy, University of Iowa,*
*Iowa City, IA, USA*

D. d'Enterria

*CERN PH/EP, CH-1211 Geneva*

CMS Collaboration

**Abstract.** The Zero Degree Calorimeter (ZDC) is integral part of the CMS experiment, especially, for heavy ion studies. The design of the ZDC includes two independent calorimeter sections: an electromagnetic section and a hadronic section. Sampling calorimeters using tungsten and quartz fibers have been chosen for the energy measurements. An overview of the ZDC is presented along with a current status of calorimeter's preparation for Day 1 of LHC.

## INTRODUCTION

The two Zero Degree Calorimeters (ZDCs) are a major hardware contribution of the US Heavy Ion group to the CMS experiment. The ZDCs will measure neutrons and very forward photons for the heavy-ion and low-luminosity (up to $10^{33}$ cm$^{-2}$s$^{-1}$) pp collisions. The design is robust, as simple as possible and based on the Relativistic Heavy Ion Collider (RHIC) ZDC design [1], which showed highly successful performance for the four RHIC BNL experiments: BRAHMS, PHOBOS, PHENIX and STAR.

The ZDCs will provide reliable measurements of real-time luminosity, beam tuning and accelerator monitoring during A+A, p+A, and p+p running. In the A+A mode, the measured coincidence rate of forward-backward neutron signals emitted in the decay of the Coulomb excited colliding ions can be used to determine the absolute A+A luminosity. The ZDCs, in addition, will help to discriminate effectively against beam-residual gas backgrounds and complement other available methods for commissioning the proton and ion beams at the LHC.

The event-by-event determination of the heavy-ion collision centrality is a basic prerequisite for any physics measurement of A+A reactions. The centrality of A+A reactions determines the degree of overlapping of the colliding nuclei and, thus, the maximum energy density reached in the interaction. Because the ZDCs are sensitive to "spectator" neutrons, they are important detectors for measurements of reaction centrality. The combination of ZDCs measuring the energy of forward neutrons, together with other multiplicity detectors at more central rapidity, has become the standard method for determining reaction centrality in A+A collisions at RHIC.

The ZDCs have been proven at RHIC to be of paramount importance for triggering and tagging UPC (ultra peripheral collision) events by measuring the forward neutrons issuing from the Coulomb-excited-nucleus dissociation.

For the photon measurement, it is necessary to have an electromagnetic zero-degree calorimeter in front of the neutron zero-degree calorimeter. The "standard" Pomeron-Pomeron diffractive A+A collision leading to central particle production and forward nucleus dissociation will not only be a background for $\gamma+\gamma$ and $\gamma+A$ processes but also a signal in their own right, accessible to study with the ZDC.

## MECHANICAL AND OPTICAL DESIGN

In order to measure neutrons and very forward photons the calorimeter will be placed downstream of the first beam dipole magnet in the straight section and between the two beam pipes. In CMS this region is located ~140 m on each side of the interaction vertex inside the neutral particle absorber (TAN), designed to protect LHC magnets by limiting their absorbed dose. This neutral particle absorber was built with a detector slot of 1 m length, 96 mm width and 607 mm height, into which are inserted 90 cm length copper absorber bars and a transversally segmented ionizing chamber (proton-proton luminosity monitor). The design of each individual ZDC includes two independent calorimeter sections, the electromagnetic section (EM section) and the hadronic section (HAD section). These sections replace the copper absorbers (see Fig. 1). The ZDC calorimeter sections are sampling calorimeters with the core of each structure consisting of a tungsten-plate/quartz-fiber-ribbon stack. A significant advantage of this technology is that the calorimeter is compact, extremely fast, and radiation hard. Quartz fibers were chosen as the active media of the ZDC calorimeters because of their unique radiation hardness features and the intrinsic speed of the Cherenkov effect. The quartz-quartz fibers can withstand up to 30 GRad with only a few percent loss in transparency in the wavelength range of 300 nm to 425 nm. The Cherenkov light generated by charged particles passing through the fibers is brought to photomultiplier tubes (PMT).

The HAD section consists of 24 layers of 15.5 mm thick tungsten plates and 24 layers of 0.7 mm diameter quartz fibers. The EM section is made of 33 layers of 2 mm thick tungsten plates and 33 layers of 0.7 mm diameter quartz fibers. The hadronic section is longitudinally segmented into 4 readout segments of ~1.4 nuclear interaction length each. The tungsten plates are tilted by 45°. The fibers of each individual tungsten/quartz fiber cell are grouped together to form readout bundles. The four fiber bundles are coupled to four photomultiplier tubes via long air core lightguides which provide optical mixing and matching of the fiber bundle to the photocathode.

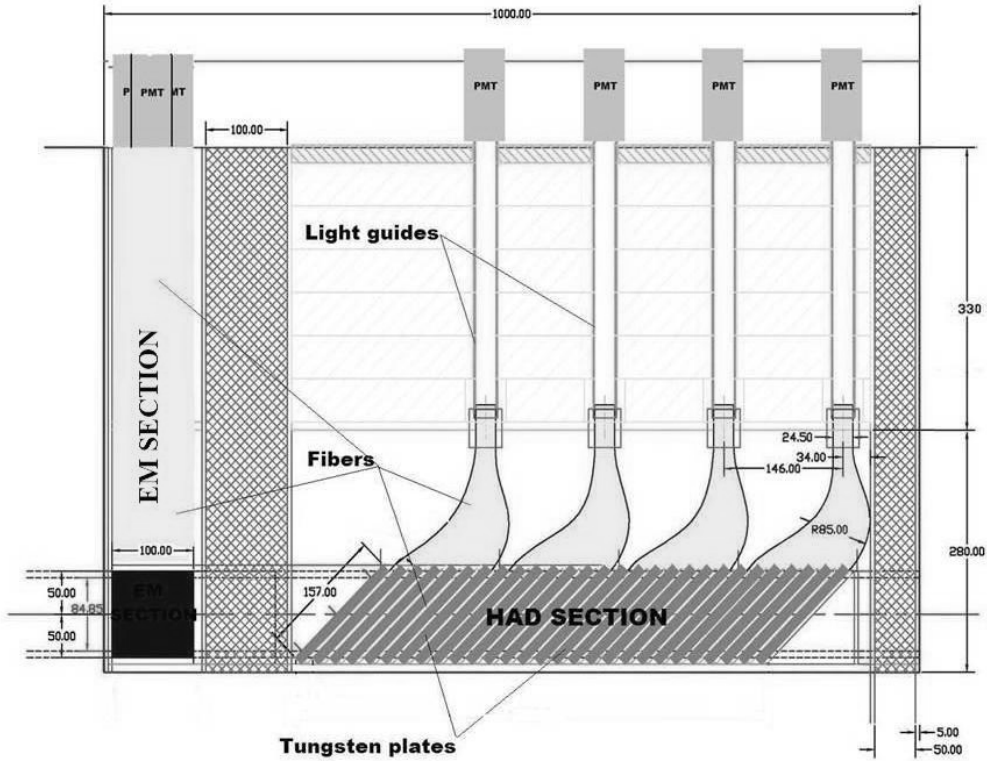

**FIGURE 1.** Schematic illustration of the TAN detector slot showing the location of the p-p luminosity monitor and the ZDC (EM section and HAD section).

The electromagnetic section is segmented into 5 horizontal (transverse) individual readout towers. The tungsten plates are oriented vertically. The fibers from each individual tower are grouped together to form a readout tower bundle. These are each coupled directly to the PMT. All calorimeter sections will be instrumented with the same type of PMT, R7525, used in the HF (forward hadronic) calorimeter. The physical characteristics of the ZDC are summarized in Table 1.

**TABLE 1. The Physical Characteristics of ZDC.**

|  | Hadronic Section | Electromagnetic Section |
|---|---|---|
| **Sampling Ratio** | 15.5 mm W/0.7 mm QF | 2 mm W/0.7 mm QF |
| **Number of Cells** | 24 | 33 |
| **Interaction (Rad.) Length** | $5.6\lambda_o$ | $(19X_o)$ |
| **Number of Channels** | 4 longitudinal segments | 5 horizontal divisions |
| **Module Size (W x L x H), mm** | 92 x 711 x 705 | 92 x 116 x 705 |
| **Weight of Module, kg** | ~400 | ~65 |

Tungsten/quartz sampling calorimeters of this type have been successfully used in a number of colliding beam and fixed target experiments, and as a consequence, the technology is well known [2]. Recent examples of detectors incorporating this technology include the zero degree calorimeters for the four heavy-ion experiments at RHIC. Much of the technology employed in the optical and mechanical design of CMS ZDC has its origins in the original research and development conducted for the ZDCs of RHIC and the HF calorimeter of the CMS experiment at the LHC, including the preparation of the optical components (quartz optical fibers, optical lightguides) and the characterization of the phototubes.

## RADIATION ENVIRONMENT

The LHC will have 7 TeV proton beams with nominal luminosity L= $10^{34}$ cm$^{-2}$s$^{-1}$ and with an interaction rate of $8\times10^8$s$^{-1}$. The power density, power dissipation, particle fluxes and spectra, accumulated dose and residual dose rates for the TAN neutral absorber have been evaluated by means of the DPMJET event generator for p-p collisions based on the MARS code [3]. During the pp runs with luminosity L= $10^{33}$ cm$^{-2}$s$^{-1}$ the expected average absorbed radiation dose will be of about 18 GRad/year. Whereas, for a one-month PbPb run the expected averaged absorbed dose will be ~30 MRad. To avoid radiation damage of the optical fibers, the ZDC will only be installed during the heavy-ion runs and the first low-luminosity p-p runs.

The phototubes will be located about ~500 mm above the beam line and ~140 m from the interaction point. According to estimates, the radiation environment will be on the same level as for the PMTs of the HF calorimeter (~10 krad/yr for the design p-p luminosity of $10^{34}$ cm$^{-2}$s$^{-1}$). This dose is permitting the use of the same Hamamatsu R7525 PMTs used by the HF calorimeter.

## PMT SYSTEM

The PMT systems of the HF calorimeter have been adopted for the ZDC. The air-core light guide of the hadronic section design is based on studies performed by the HF group at the University of Iowa. These studies show that the diffusion of light due to the mixer is optimum. Thus the 20% variation in quantum efficiency that might be expected as a result of the non-uniform response of various fibers is reduced to a few percent as a result of utilization of the mixer. The PMTs for the ZDC are eight-stage Hamamatsu R7525 phototubes with bi-alkali photocathodes, resulting in an average quantum efficiency for Cherenkov light of about 10%. The resistive chain high voltage base (ratio "B") is optimized to achieve a high gain and the maximum dynamic range of linearity (linearity is on the level of 2% for dynamic range of up to $10^3$). The PMT and base are housed in a shielding enclosure (housing). All housing components, except for the μ-metal shields, are fabricated in the University of Kansas (KU) physics machine shop.

# PROTOTYPING

An engineering mechanical prototype was designed and constructed at KU. The test module design was virtually identical to one quarter of the HAD section design (one longitudinal tower). The module was used to establish assembling protocol and to verify the assembly tolerances and clearances. Fiber routing and fiber ribbons preparation was also studied with the first mechanical prototype. Experience gained through these studies was important in establishing the final fiber routing scheme and fiber ribbon design.

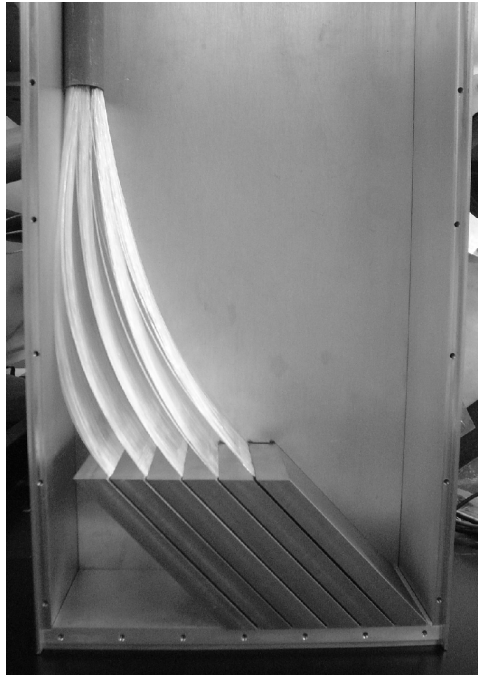

**FIGURE 2.** Picture showing the mechanical prototype of one longitudinal tower of the HAD section.

In consultation with the LHC integration group, many of the details of how to install the ZDC were worked out this year. A dummy ZDC was made from plastic and successfully inserted into the two TANs that will be used for CMS. Finally a wooden full scale copy of a TAN has been constructed at CERN to help sort out installation issues on the surface.

# PERFORMANCE

The ZDC sections have been studied with both stand-alone simulations and CMS GEANT4 simulations. These studies have been directed both towards optimization of the calorimeter design and towards understanding the detector performance. Following the simulation studies, we have fixed the main version of the converter for the EM section to be 33 tungsten plates with a thickness of 2 mm ($\sim 19 X_o$) and quartz fibers with a core diameter of 0.6 mm. We required that the energy resolution should be better than 10% for 50 GeV photons and that the calorimeter should be linear within 2% in the energy range from 10 GeV to 100 GeV. The simulation shows that a calorimeter with this geometry meets our requirements. The resulting energy resolution is displayed in Fig. 3.

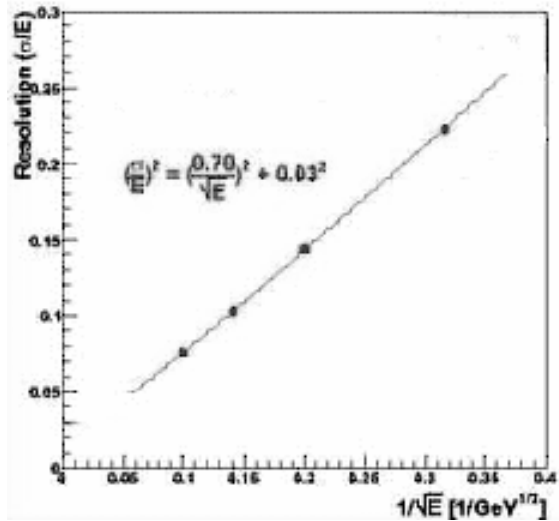

**FIGURE 3.** Energy resolution of the ZDC electromagnetic section.

The transverse cell size (in the horizontal plane) for the electromagnetic calorimeter is fixed at ~18 mm. This ensures that the main energy deposition from a single particle occurs in one cell. The 50 GeV photon shower profile is given in Fig. 4. The width of the shower is ~5 mm and the separation of two particles with a relative distance of about 2 cm in the horizontal plane at the entrance to the calorimeter is easy to distinguish. The standard technique of shower-profile fitting can be applied to separate closer showers. In practice, this position resolution will allow measurement of the beam crossing angle with a resolution of ~10 mrad.

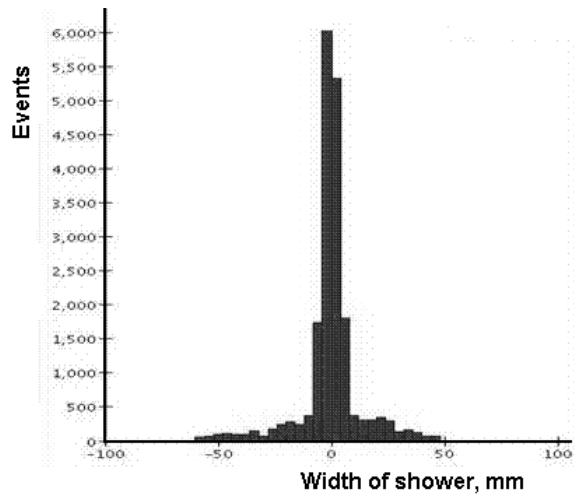

**FIGURE 4.** The 50 GeV photon shower profile in the EM section of ZDC.

The ZDC hadronic section will be in the very forward region, and thus the calorimeter needs to be quite dense so as to absorb hadrons with an energy of a few TeV. The choice of tungsten as the passive material meets this requirement. We have studied different tungsten plate thickness varying from 5 mm to 20 mm and we did not find significant changes in energy resolution for energies in the region from 500 GeV to 3 TeV. For quartz fibers with core diameter of 0.6 mm the energy resolution will be in the range 10% ± 2%, which is better than our requirement of 15% for 3 TeV. For practical reasons, such as the number of tungsten plates, the number of fiber ribbons, and the limited space, we chose 15.5 mm tungsten plates as the absorber in the HAD section. The energy resolution for a calorimeter with 10 mm-thick tungsten plates is presented in Fig. 5.

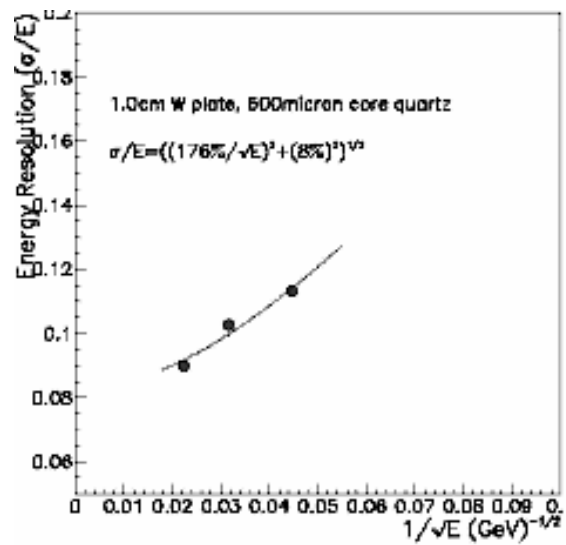

**FIGURE 5.** The energy resolution of the hadronic section of ZDC.

## ELECTRONICS

The ZDC electronics includes the four major functions necessary for the functioning and data reading from the detector: front-end electronics, trigger electronics, data acquisition, and the high voltage system needed for operation of the phototubes.

In the nucleus-nucleus running mode, the ZDC is a central detector for Level-1 triggering. The standard ZDC minimum bias trigger uses a coincidence of the signals from the two calorimeters (right and left) on both sides of the interaction point. In addition, a single (or double) signal from one (or both) ZDCs will be used to trigger on ultraperipheral electromagnetic nucleus-nucleus interactions leading to the dissociation of one (or both) colliding nuclei. These coincidences are processed by the ZDC electronics in the counting room.

The front-end electronics, including signal processing, digitization, buffering,

formation of an event trigger, and the first level of readout, is located in HCAL VME crates as close to the ZDC as possible. Because of the time delays introduced these crates located in the adjacent underground counting room (USC55).

There are a total of 18 readout channels for the two ZDCs. The signals from the ZDCs are transmitted through a long (210 m) coaxial cable to the counting room. Signals from the EM channels will go directly to the QIE ASIC. The signal from each HAD channel will be split into two signals; one signal will go to the QIE and the other will be summed with the other three HAD channel signals. The analog sum of the four signals related to each HAD section, proportional to the total energy deposition in each detector and proportional to the total number of spectators coming out from the interaction, will be sent to the splitter. One output signal of the splitter will go to the QIE to provide a Level 1 trigger for several centrality intervals, and other signal will go to the input of the discriminator and then to a logic unit for coincidence. The coincidence of two signals from both sides of the interaction point is sensitive to most of the nuclear and electromagnetic cross section. Information from scalers will be used for tuning the interaction of beams and for defining the real-time luminosity.

## INSTALLATION AND PLANS

The ZDC detector configuration necessary to accomplish all of the physics goals includes 2 ZDCs (two hadron sections and two electromagnetic sections). The Day 1 detector implementation will provide physics results. The installation operation is scheduled to start early September of 2006. After all connections are complete, each section will be tested and commissioned. Testing and commissioning of the detectors will mainly involve HV (high voltage) system tests and LED (light emitting diode) and laser calibrations. Each ZDC section will have been calibrated in the test beam. We foresee all functionality of the read-out to be tested and operational shortly after the cables are connected and the electronics in the counting room is functioning.

## ACKNOWLEDGMENTS

We thank S. J. Sanders for useful discussions and support. D. Macina, E. Tsesmelis and A.-L. Perrot of CERN TS/LEA provided invaluable help for the integration and installation of the ZDC. D. d'Enterria acknowledges support by the 6th EU Framework Programme (contract MEIF-CT-2005-025073). We are grateful to the entire US Heavy Ion group for support. This project was supported by the HENP Divisions of the U.S.DOE and the U.S.NSF.

## REFERENCES

1. C. Adler et al., *Phys. Rev. Lett* **89**, 272302 (2002); A. Denisov et al., *Nucl. Phys.* **A698,** 551 (2002)**.**
2. Y. Onel, 10[th] *International Conference on Calorimetry in High Energy Physics* (CALOR 2002), Pasadena, CA, 25-30 March 2002; N. Akchurin and R. Wigmans, *Rev.Sci.Instrum* **74,** 2955-2972 (2003).
3. N. V. Mokhov et al., FERMILAB-FN-732.